\documentclass[conference]{IEEEtran}
\usepackage{amsmath,amssymb}
\usepackage{color}
\usepackage[pdftex]{graphicx}
\usepackage{mathtools}
\usepackage{fancyvrb}
\usepackage{lineno,hyperref}
\usepackage{multirow}
\usepackage{fancyvrb}
\onecolumn

\newcommand{\qed}{\nobreak \ifvmode \relax \else
      \ifdim\lastskip<1.5em \hskip-\lastskip
      \hskip1.5em plus0em minus0.5em \fi \nobreak
      \vrule height0.75em width0.5em depth0.25em\fi}

\begin{document}

\title{Fast Shortest Path Routing in Transportation Networks with Time-Dependent Road Speeds}
\author{\IEEEauthorblockN{Costas K. Constantinou\IEEEauthorrefmark{1},
Georgios Ellinas\IEEEauthorrefmark{1}\IEEEauthorrefmark{2}, 
Christos Panayiotou\IEEEauthorrefmark{1}\IEEEauthorrefmark{2}
and
Marios Polycarpou\IEEEauthorrefmark{1}\IEEEauthorrefmark{2}}
\IEEEauthorblockA{\IEEEauthorrefmark{1}KIOS Research Center for 
Intelligent Systems and Networks}
\IEEEauthorblockA{\IEEEauthorrefmark{2}Department of Electrical and Computer Engineering\\
University of Cyprus\\
Emails: \{constantinou.k.costas, gellinas, christosp, mpolycar\}@ucy.ac.cy}}
\maketitle

\begin{abstract}
The current paper deals with the subject of shortest path routing in transportation networks (in terms of travelling time), where the speed in several of the network's roads is a function of the time interval. The main contribution of the paper is a procedure that is faster compared to the conventional approaches, that derives the road's traversal time according to the time instant of departure, for the case where the road's speed has a constant value inside each time interval (in general, different value for each time interval). Furthermore, the case where the road's speed is a linear function of time inside each time interval (in general, different linear function for each time interval) is investigated. A procedure that derives the road's traversal time according to the time instant of departure is proposed for this case as well. The proposed procedures are combined with Dijkstra's algorithm and the resulting algorithms, that are practically applicable and of low complexity, provide optimal shortest path routing in the networks under investigation.    
\end{abstract}

\section{Introduction}

In the current paper, the subject of shortest path routing in transportation networks where the roads' speed is function of time, is investigated. This category of networks is met in practice, since the speed in network's roads can be available for several time instants, either directly measured, or derived from vehicle density measurements \cite{may}.

The objective of this work was the development of low-complexity, practically applicable algorithms for fast derivation of shortest paths (in terms of travelling time) in the networks under investigation.      

Section \ref{def} gives the notation, definitions and assumptions used throughout the paper. In general, a transportation network is modelled as a directed graph, where the nodes, arcs of the graph represent the network junctions, roads respectively. A cost is assigned to each arc, that in the current work represents the travelling time on the arc (called ``traversal time'' as well). For the classical case, where the network is \emph{time-independent} (called \emph{static} as well), this cost is constant over time for every arc, derived by dividing the length of the arc with the speed in it.  For the case investigated here, the network is \emph{time-dependent}, i.e., this cost is a function of time, depending on the time instant of departure. Therefore, for shortest path routing in time-dependent networks an additional calculation must be performed compared to static networks: the calculation of the travelling time on the network's arcs according to the time instant of departure. 

Throughout the paper it is considered that the time horizon is split into non-overlapping time intervals, and the speed in a network's arc depends on the time interval. This model was firstly presented in \cite{Sung}. More details on this, as well as the reasons why this model was adopted for the work proposed here, can be found in Section \ref{models}, where the main models for the networks under investigation, are presented.  

The main contribution of the current paper can be found in Sections \ref{routing} and~\ref{general}. 

In Section \ref{routing}, a procedure is proposed, that derives the arc's (road's) traversal time according to the time instant of departure, for the case where the speed in the arc has a constant value inside each time interval (in general, a different value for each time interval). The proposed procedure is faster compared to the conventional approaches. More precisely, if we consider that the time horizon is split into $K$ time slots, the computational complexity of the proposed procedure is of order $O(\log K)$, whereas the complexity of the fastest existing approach is of order $O(K)$. Special cases are also investigated, where the complexity of the proposed procedure can be further decreased if certain constraints are valid for the network graph. 

In Section \ref{general}, the general case of speed being an arbitrary function of time \emph{inside} the time interval is investigated, and a procedure is proposed that derives the road's traversal time according to the time instant of departure, for  the case where the speed in the arc is linear function of time inside each time interval (in general, a different linear function for each time interval). To the best of our knowledge, relevant methods for this case do not exist in the literature.  

The proposed methods are combined with Dijkstra's algorithm and the resulting algorithms, that are practically applicable and of low complexity, provide optimal shortest path routing in the networks under investigation.

The paper finishes with Section \ref{conclusions}, where the conclusions are presented, as well as possible future research.

\section{Notation, Definitions,  Assumptions} \label{def}

Throughout the paper the network is modelled as a directed graph $G=(N, A)$, consisting of $n=|N|$ nodes (road junctions) and $m=|A|$ arcs (roads). The arc originating from node $x$ and end at node $y$ is denoted by $<xy>$,  and its length (i.e., actual length of the corresponding road) by $d^{xy}$. The speed in $<xy>$ is denoted by $v^{xy}$. A node $y$ is considered to be adjacent to $x$ if $<xy>$ exists in the graph.

The cost $c^{xy}$ (or ``traversal time'' or ``travelling time'') of arc $<xy>$ is defined as the time needed to traverse it, i.e., to move from node $x$ to $y$. Consequently, the term ``shortest path'' from node $x$ to node $y$ refers to the path of the minimum travelling time from $x$ to $y$. For the case of static networks, the cost of an arc is constant over time, equal to $d^{xy}/v^{xy}$ for arc $<xy>$. For time-dependent networks, it is a function of the time instant of departure ($\tau$) from node $x$, and it is denoted by $c^{xy}(\tau)$. Therefore, for shortest path routing in time-dependent networks an additional calculation must be performed compared to static networks: the calculation of $c^{xy}(\tau)$ according to the time instant of departure from $x$. 

Throughout the paper, it is considered that the speed in the network's roads has been measured for certain time instants\footnote{Either directly measured, or derived from vehicle density measurements \cite{may}}, during a large time interval (e.g., an entire year). In this way, a speed-pattern can be derived for each road. These patterns constitute an estimation of the time-dependent network graph.

\section{Existing Models of Time-Dependent Transportation Networks} \label{models}

In this Section, the two main models of time-dependent transportation networks are presented.

\subsection{Flow Speed Model (FSM)} \label{fsm}

One approach to model a time-dependent network is to use the \emph{Flow Speed Model (FSM)}, proposed in \cite{Sung}. In the FSM, for each network arc a temporal area from $t=0$ to $t=T$ is partitioned into (in general not equal) non-overlapped time intervals, and the speed depends on the time interval. The time division is the same for every network arc. The $(k+1)^{th}$ time interval is denoted by $[\tau_k, \tau_{k+1})$, with  $k \in \mathcal{N}$ and $0\leq k \leq K-1$, $\tau_0=0$ and $\tau_K=T$. The number of time intervals is equal to $K$.  The speed in arbitrary arc $<xy>$ is considered to be constant inside each time interval and it is denoted by $v^{xy}_{k}$ for time interval $[\tau_k, \tau_{k+1})$. Let the set of speeds $v^{xy}_{k}$ and time intervals $[\tau_k, \tau_{k+1})$, $0\leq k \leq K-1$, for arc $<xy>$ be denoted by $V^{xy}$ and $T^{xy}$ respectively, and $V=\cup_{\forall <xy>\in A} V^{xy}$ and $T=\cup_{\forall <xy>\in A} T^{xy}$. Then, for the FSM the network graph is denoted by  $G=(N, A, T, V)$. 

For $t>T$, it can be considered either that $v_k^{xy}=v_{K-1}^{xy}$ (i.e., the network is static for $t>T$), or  that $v_k^{xy}$ is periodic with period equal to $T$.

The FSM always satisfies the First-In-First-Out (FIFO) property, as proven in \cite{Sung}. In simple words, the FIFO property has the following meaning: Consider two vehicles $A$ and $B$ that depart from node $x$ and traverse arc $<xy>$, with $B$ having greater time instant of departure from $x$, compared to $A$. Then, $B$ will have greater time instant of arrival at $y$,  compared to $A$.

As stated in Section \ref{def}, for routing purposes in time-dependent networks, the derivation of arc traversal time must be performed. In the FSM, this is performed as follows. 

\subsubsection{Procedure for Derivation of Arc Traversal Time ($ATT$)} \label{Derivation}
The traversal time $c^{xy}(\tau)$ of arc $<xy>$ if $\tau$ is the time instant of departure from $x$ can be derived as follows ($ATT^{xy}(\tau)$ procedure\footnote{This procedure was initially proposed in \cite{Sung}; the one presented in the current paper is an equivalent form of the one in \cite{Sung}.}):

\begin{enumerate}
    \item Locate index $k$ such that $\tau_k \leq \tau < \tau_{k+1}$ \label{attindex}
    \item If ($v_k^{xy}\cdot(\tau_{k+1}-\tau)\geq d^{xy}$) $c^{xy}(\tau)=\frac{d^{xy}}{v_k^{xy}}$

    Else $\{$
    \item
    \begin{enumerate}
        \item 
        \begin{enumerate}
            \item $a=d^{xy}-v_{k}^{xy}\cdot(\tau_{{k}+1}-\tau)$
            \item $k^*=k+1$
        \end{enumerate}
        \item While ($v_{k^*}^{xy}\cdot(\tau_{k^*+1}-\tau_{k^*})<a$) $\{$  \label{attwhile}
        \begin{enumerate}
            \item $a\leftarrow a-v_{k^*}^{xy}\cdot(\tau_{{k^*}+1}-\tau_{k^*})$
            \item ${k^*}\leftarrow {k^*}+1$ $\}$ \label{stepk}
        \end{enumerate}
    \end{enumerate}
    \item $c^{xy}(\tau)=(\tau_{{k^*}}-\tau)+\frac{a}{v_{k^*}^{xy}}$  $\}$
\end{enumerate}

The necessity of the aforementioned procedure is due to the fact that a single time interval may not  be enough for the derivation of the arc traversal time. This occurs when the distance that can be traversed from the time instant of departure till the end of the corresponding time interval, is less than the length of the arc.

Step \ref{attindex} of the $ATT$ procedure needs $O(K)$ time, if the time intervals are checked sequentially. The order of the number of the time intervals that are checked during the while loop (step \ref{attwhile}) is $O(K)$. Therefore, the order of the computational complexity of the $ATT$ procedure is $O(K)$. 

\begin{figure}
\centering{\includegraphics[scale=0.2]{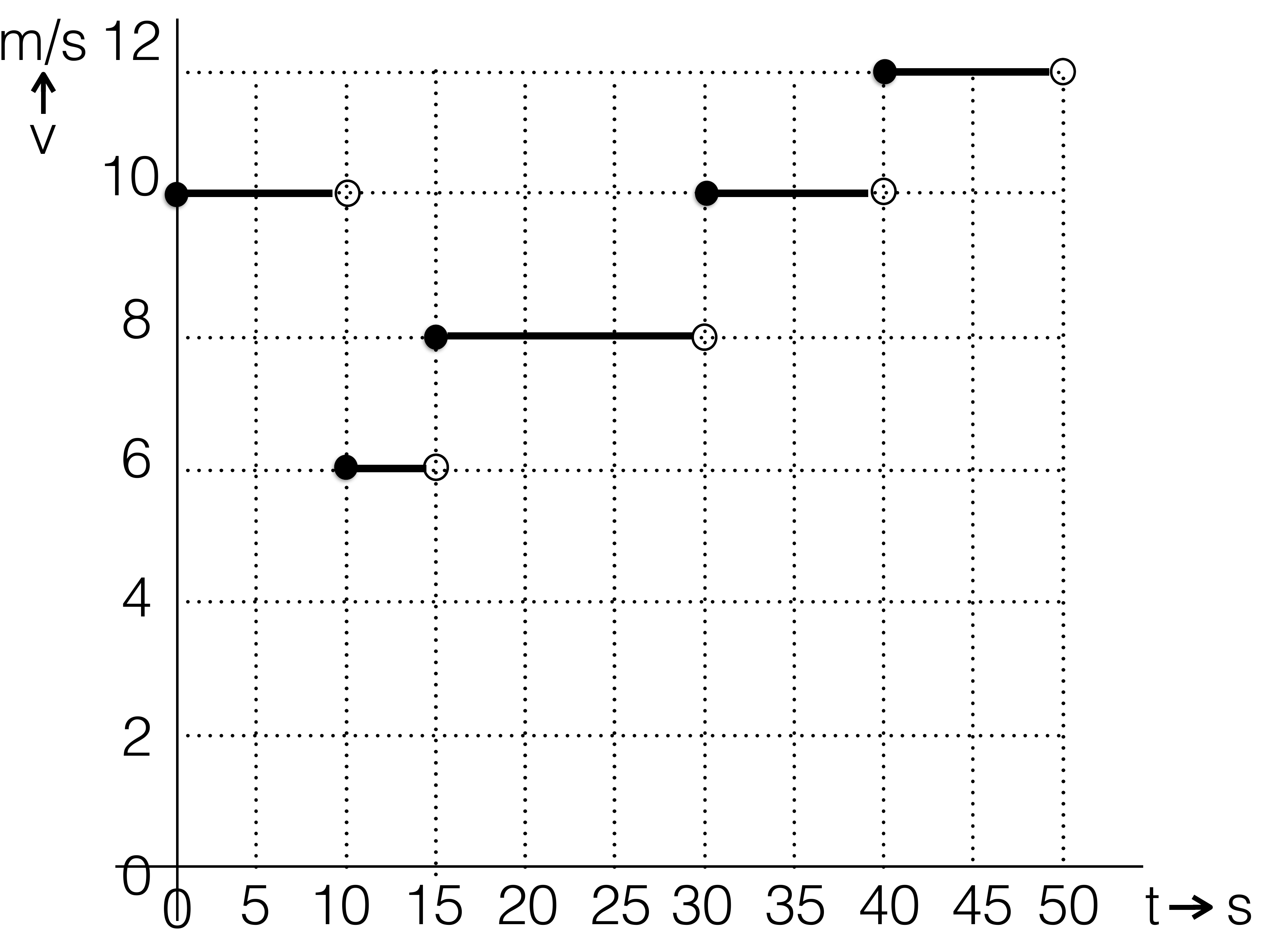}}
\caption{Example of derivation of arc traversal time ($d=250m$, $\tau=2s$)} \label{example}
\end{figure}

The example of Figure \ref{example} illustrates the operation of the $ATT$ procedure. Here, the length $d^{xy}$ of arc $<xy>$ is equal to $170m$ and the departure time $\tau$ from $x$ is equal to $6s$. The first five time intervals are $[\tau_0, \tau_1)=[0,10)$,  $[\tau_1, \tau_2)=[10,15)$,  $[\tau_2, \tau_3)=[15,30)$,  $[\tau_3, \tau_4)=[30,40)$, $[\tau_4, \tau_5)=[40,50)$. These along with their corresponding speed are given in Figure~\ref{example}. For this example, the $ATT$ procedure performs as follows:

\begin{enumerate}
    \item The departure time $\tau=6s$ lies into time interval $[\tau_0, \tau_1)=[0,10)$ $\Rightarrow$ $k=0$
    \item The distance that can be traversed from $\tau=6s$ until the end of this time interval is equal to $v_0^{xy}\cdot(\tau_1-\tau)=10\cdot(10-6)=40m$. Since it is smaller than the length of the arc, the procedure is continued.
    \item
    \begin{enumerate}
        \item 
        \begin{enumerate}
            \item $a=d^{xy}-v_{0}^{xy}\cdot(\tau_{1}-\tau)=170-40=130m$
            \item ${k^*}=1$
        \end{enumerate}       
        \item 
        \begin{itemize}
            \item $v_{1}^{xy}\cdot(\tau_{2}-\tau_{1})=6\cdot(15-10)=30m<a=130m \Rightarrow$
            \begin{enumerate}
                \item $a=130-30=100m$
                \item ${k^*}=2$ 
            \end{enumerate}
            \item $v_{2}^{xy}\cdot(\tau_{3}-\tau_{2})=8\cdot(30-15)=120m>a=100m \Rightarrow$ exit from the while loop
        \end{itemize}
    \end{enumerate}
    \item $c^{xy}(6)=(\tau_2-\tau)+a/v_2^{xy}=(15-6)+100/8=9+12.5=21.5s$
\end{enumerate}

The traversal time is equal to $21.5s$ and the time instant of arrival at node $y$ is equal to $6+21.5=27.5s$.

\subsubsection{Applicability of the FSM for Networks under Investigation}

The application of the FSM for the networks investigated in the current paper, is straightforward. The time instants for partitioning the temporal area are the ones where the speed has been measured. As stated previously, in \cite{Sung} the time division is the same for every network arc. In general, it may be different. In the current paper, for simplicity, we make the same assumption as in \cite{Sung}, i.e., that the time division is the same for every arc. Nevertheless, the generalisation is straightforward, as can be seen from the description of the $ATT$ procedure. 

In \cite{Sung}, as stated previously, the speed is considered to be constant in every time interval. Up to Section \ref{routing}, this assumption is adopted. The general case of the speed being a function of time \emph{inside} a time interval, is investigated in Section~\ref{general}. 

In \cite{Sung}, the $ATT$ procedure is combined with Dijkstra's algorithm \cite{dijkstra}, and the resulting algorithm  (called \emph{Time-Dependent-Dijkstra (TD-Dijkstra)} hereafter) is as follows.

\subsubsection{Time-Dependent Dijkstra's Algorithm}

As stated in Section~\ref{def}, for the networks under investigation, the calculation of $c^{xy}(\tau)$ must be performed during the execution of any shortest path routing algorithm. In \cite{Sung}, where the FSM model and the $ATT$ procedure were proposed, the latter is combined with Dijkstra's algorithm \cite{dijkstra} and the resulting algorithm (TD-Dijkstra) is suitable for shortest path routing in networks with time-dependent road speeds. The input of TD-Dijkstra is the network graph $G=(N, A, T, V)$ and the source node, and the output is the shortest path from the source to every other network node. For the execution of TD-Dijkstra, the following are used.

\begin{itemize}
    \item $s$: Source.
    \item $W(x)$: Label of node $x$. 
    \item $p(x)$: Predecessor of node $x$. 
    \item $G_x$: Set of nodes adjacent to node $x$.
    \item $g_x$: Number of nodes adjacent to node $x$ (i.e., $g_x=|G_x|$).
\end{itemize}

The exact steps of TD-Dijkstra are: 
\begin{enumerate}
    \item \label{step1}
    \begin{enumerate}
        \item $W(s)=0$
        \item $p(s)=0$
        \item $N^*=N-\{s\}$
        \item $ \forall x\in N^*$: 
        \begin{enumerate}
            \item If ($x\in G_s$) 
            
            $\{$ Run $ATT^{sx}(W(s))$; $W(x)=c^{sx}(W(s))$; $p(x)=s$ $\}$ \label{stepa}
            \item Else $\{$ $W(x)=\infty$; $p(x)=0$ $\}$
        \end{enumerate}    
    \end{enumerate}
    \item While ($N^*\neq \emptyset$)$\{$ \label{step2}
    \begin{enumerate}
        \item Find $x\in N^*$ such that $\forall x'\in N^*$: $W(x)\leq W(x')$ 
        \item $N^*\leftarrow N^*-\{x\}$
        \item $\forall x'\in (N^*\cap G_x)$: 
         \begin{enumerate}
            \item Run $ATT^{xx'}(W(x))$ \label{stepb}
            \item If ($W(x)+c^{xx'}(W(x))<W(x')$) 
            
            $\{$ $W(x')=W(x)+c^{xx'}(W(x))$; $p(x')=x$ $\}$ $\}$ 
         \end{enumerate}   
    \end{enumerate}
\end{enumerate}

The TD-Dijkstra algorithm functions as the classical Dijkstra's algorithm, with the difference that the $ATT$ procedure is used in steps \ref{stepa} and \ref{stepb} for the derivation of the arc's cost according to the time instant of departure.   

On termination of the algorithm, the label $W(x)$ of a node $x$  gives the cost of the shortest path from the source to this node, and $p(x)$ gives its predecessor in this path.

Since the computational complexity of the $ATT$ procedure is of order $O(K)$, step \ref{step1} requires $O(nK)$ time and each iteration of step \ref{step2} requires $O(n+g_xK)$ time. Therefore, due to the fact that $\sum_{x=1}^{n}g_x=m$,  the complexity of TD-Dijkstra is of order  $O(nK+n^2+mK)=O(n^2+mK)$.

The D-Dijkstra is, to the best of our knowledge, the fastest existing algorithm for shortest path routing in the networks investigated in the current paper.

\subsection{Traversal Time Model (TTM)} \label{ttm} 

Another approach to model a time-dependent network, is to use the \emph{Traversal Time Model (TTM)}. In the TTM, the \emph{traversal time function} $f^{xy}(\tau)$ is utilised, where $f^{xy}(\tau)$ is equal to the time needed to traverse arc $<xy>$, if $\tau$ is the time instant of departure from node $x$. The TTM may or may not satisfy the FIFO property. This model was initially proposed in \cite{Cooke} and was exploited in several other papers, such as in \cite{Delling}, \cite{Nannicini}, \cite{Delling2}, \cite{Delling3}, \cite{Delling4}, \cite{Ding}, \cite{Batz}, \cite{Chabini}.


Function $f^{xy}(\tau)$ can be defined so as to have either integer-valued or real-valued domain and range, leading to discrete- or continuous-time-dependent networks respectively.

If $f^{xy}(\tau)$ is defined so as to have integer-valued domain and range (first variation of the TTM), the network is discrete-time and $f^{xy}(\tau)$ is known for every $\tau \in \mathcal{N}$, $0\leq \tau \leq T$. This variation can be found in \cite{Chabini} as well as other sources.

For real-valued domain and range of $f^{xy}(\tau)$ (second variation of the TTM), the network is continuous-time and $f^{xy}(\tau)$ is known for several time instants denoted by $\tau_k$, with $0\leq \tau_k \leq T$, $k \in \mathcal{N}$ and $0 \leq k \leq K$, $\tau_0=0$ and $\tau_K=T$.

These time instants, in general, are different for each arc. Work that utilises this variation of the TTM can be found in \cite{Delling}-\cite{Batz} and possibly in other sources. In the aforementioned papers, $f^{xy}(\tau)$ is considered to be a \emph{piecewise linear function of time}, having the time instants $\tau_k$ where it is known, as breakpoints. For a time instant $\tau$ for which $f^{xy}(\tau)$ is unknown, it is derived by linear interpolation between the consecutive breakpoints $\tau_{k}$, $\tau_{k+1}$ such that $\tau_{k}<\tau<\tau_{k+1}$:

\begin{eqnarray}
&\frac{f^{xy}(\tau)-f^{xy}(\tau_k)}{\tau-\tau_k}=\frac{f^{xy}(\tau_{k+1})-f^{xy}(\tau_k)}{\tau_{k+1}-\tau_k}&  \label{eqnA} \\
\Rightarrow &f^{xy}(\tau)=\frac{f^{xy}(\tau_{k+1})-f^{xy}(\tau_k)}{\tau_{k+1}-\tau_k}\cdot (\tau-\tau_k)+f^{xy}(\tau_k)& \label{eqnB}
\end{eqnarray}

\begin{figure}
\centering{\includegraphics[scale=0.25]{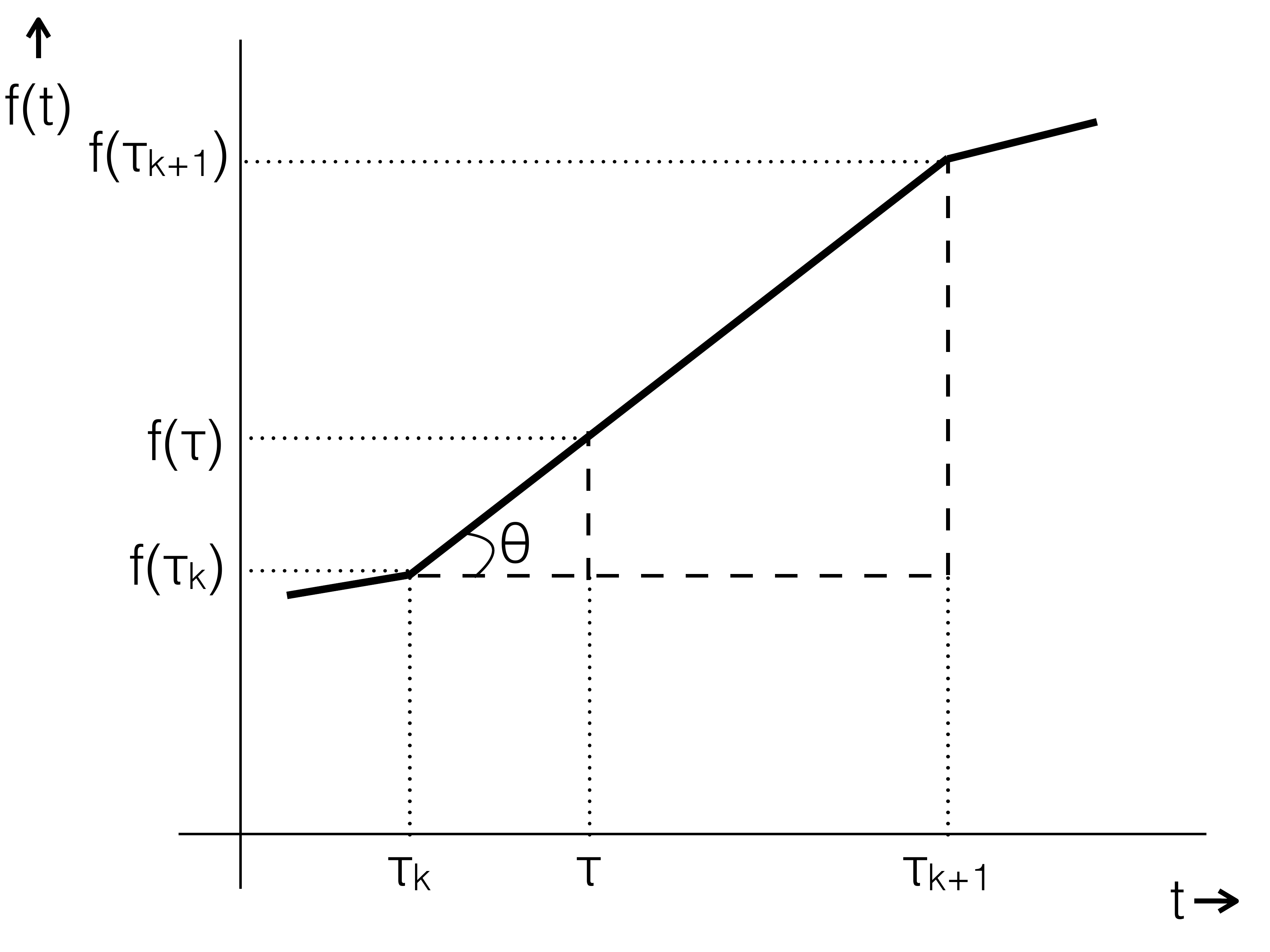}}
\caption{Derivation of equation \ref{eqnA}} \label{fig2}
\end{figure}

Equation \ref{eqnA} can be derived from the tangent of angle $\theta$ of Figure \ref{fig2}. 

For both variations,  for $\tau>T$ either it can be considered that $f^{xy}(\tau)=f^{xy}(T)$ (i.e., the network is static for $\tau>T$), or it can be considered that $f^{xy}(\tau)$ is periodic with period equal to $T$. 

\subsubsection{Applicability of the TTM for Networks Under Investigation} \label{nonlinear}

For optimal routing in the networks under investigation the continuous-time variation of the TTM must be used, since the time instant of departure from a network node can have any arbitrary value.  The derivation of $f^{xy}(\tau)$ for the time instants the speed has been measured, can be performed using the $ATT$ procedure of the FSM. For the calculation of $c^{xy}(\tau')$ for $\tau'$ not equal to one of the aforementioned time instants, the  assumption that $f^{xy}(\tau)$ is piecewise linear, must be valid. However, the following example shows that for the networks investigated in the current paper, this is not always true.

Consider the example of Figure~\ref{example}. Here, if the $ATT$ procedure is applied for $\tau$ equal to $0s$, $10s$, the derived $f^{xy}(\tau)$ is equal to $20s$, $22s$, respectively. Using equation \ref{eqnB}, $f^{xy}(6)=21.2s$. However, $c^{xy}(6)=21.5s \neq f^{xy}(6)$, as derived in Section \ref{Derivation}. Therefore, for this example, the assumption that  $f^{xy}(\tau)$ is piecewise linear, is not valid. 

Consequently, the FSM is utilised for the routing algorithms proposed in the following Section.

\section{Fast Shortest Path Routing Algorithms} \label{routing}

The design of a procedure that functions as the $ATT$\footnote{i.e., that derives the cost of an arc according to the time instant of departure} but with lower computational complexity compared to the $ATT$, can lead to algorithms for faster shortest path routing in the networks under investigation, compared to TD-Dijkstra. 

As stated in Section \ref{Derivation}, the complexity of the $ATT$ procedure is of order $O(K)$ due to steps \ref{attindex} and \ref{attwhile}. When the $ATT$ procedure is executed, the index of the time interval where the time instant of arrival lies in, can be kept \cite{Sung}. In more detail, after the execution of the $ATT$ for the derivation of the time instant $\tau^*=\tau+c^{xy}(\tau)$ of arrival at node $y$ of the arbitrary arc $<xy>$, the index $k^*$ such that $t_{k^*}\leq \tau^* <t_{k^*+1}$ can be kept, since it is known from the last iteration of step \ref{stepk} of the $ATT$ (let the index that is kept, be denoted by $I$). For each one of the upcoming executions of the $ATT$ in step \ref{stepb} of TD-Dijkstra, performed for every $z$ such that $z\in (N^*\cap G_y)$, index $k$ at step \ref{attindex} is equal to $I$. Therefore,

\begin{itemize}
    \item For the first execution of the $ATT$ (step \ref{stepa} of TD-Dijkstra) the index $k$ of step \ref{attindex} of the $ATT$ is known (it is equal to $0$), and
    \item For any upcoming execution of the $ATT$ (step \ref{stepb} of TD-Dijkstra) the index $k$ of step \ref{attindex} of the $ATT$ is known as described previously
\end{itemize}

The result is that the complexity of step \ref{attindex} of the $ATT$ is of order $O(1)$ if for every execution of the $ATT$, the index of the time interval where the time instant of arrival lies in, is kept.

Consequently, any attempt to decrease the complexity of the $ATT$ procedure must concentrate on the reduction of the complexity of step \ref{attwhile} of it. At this step, variable $k^*$ is increased by one at each iteration, (sequential search is performed over $k^*$) until the last necessary time interval is located, i.e., the time interval where the time instant of arrival $\tau^*$ at node $y$ (of $<xy>$), lies in (i.e., $[t_{k^*}, t_{{k^*}+1})$ such that $t_{k^*}\leq \tau^*<t_{{k^*}+1}$). 

An initial thought for decreasing the complexity would be to perform binary search \cite{ahuja} over variable $k^*$  instead of sequential search, to locate the last necessary time interval. However, this will not lead to lower complexity, since at each iteration of the binary search, the check whether $t_{k^*}\leq \tau^*<t_{{k^*}+1}$ will need $O(K)$ time\footnote{Since at each iteration of this check,  the initial while loop, i.e.,  with sequential search on $k^*$, must be executed.}, thus not decreasing the complexity of the $ATT$ procedure. 

To achieve reduction of the complexity of the $ATT$ procedure using binary search over $k$, a pair of new variables, named \emph{effective length} and \emph{additive effective length}, are introduced in the current paper. They are defined as follows.

\subsection{Effective Length, Additive Effective Length}

\begin{itemize}
    \item The \emph{effective length} $l_k^{xy}$ is defined as the distance that can be traversed on arc $<xy>$ during the time interval $[t_k, t_{k+1})$. Therefore, 
  
\begin{equation}
    l_k^{xy}=v_k^{xy}\cdot(t_{k+1}-t_k) \label{effective1}
\end{equation} 

Note that in general, $l_k^{xy}$ can be greater, equal or less than the actual length $d^{xy}$ of arc $<xy>$.

    \item The \emph{additive effective length} $L_i^{xy}$ of the time interval $[t_i, t_{i+1})$ for arc~$<xy>$ is defined as $L_i^{xy}=\sum_{k=0}^{i}l_k^{xy}$, i.e., is the sum of the effective lengths from the first time interval $[t_0, t_{i+1})$ up to the $[t_i, t_{i+1})$, including the latter. In other words, it is equal to the distance that is traversed from time instant $t_0$ to $t_{i+1}$. For $i\leq j$,
    
\begin{equation}
L_j^{xy}-L_i^{xy}=\sum_{k=0}^{j}l_k^{xy}-\sum_{k=0}^{i}l_k^{xy}=\sum_{k=i+1}^{j}l_k^{xy} \label{eqnld}
\end{equation}

Therefore, the difference $L_j^{xy}-L_i^{xy}$ is equal to the distance that is traversed from time instant $t_{i+1}$ to $t_{j+1}$.  
\end{itemize}

In the following Section, a new procedure is proposed, named \emph{Fast $ATT$ procedure ($FATT$)}, that utilises the additive effective length for fast calculation of the arc traversal time. 

\subsection{Fast $ATT$ Procedure ($FATT$)} \label{fattc}

The exact steps of the $FATT^{xy}(\tau)$ procedure for arc $<xy>$ and $\tau$ as time instant of departure from node $x$, are as follows.

\subsubsection{Steps of the $FATT^{xy}(\tau)$ Procedure}

\begin{enumerate}
    \item Locate index $k$ such that $\tau_k \leq \tau < \tau_{k+1}$ \label{fattindex}
    \item If ($v_k^{xy}\cdot(\tau_{k+1}-\tau)\geq d^{xy}$) $c^{xy}(\tau)=\frac{d^{xy}}{v_k^{xy}}$

    Else $\{$
    \item
    \begin{enumerate}
        \item 
        \begin{enumerate}
            \item $a=d^{xy}-v_k^{xy}\cdot(\tau_{k+1}-\tau)$
            \item $k^*= \lfloor \frac{(k+1)+K}{2} \rfloor$
            \item $a'=0$
        \end{enumerate} 
        \item While ($a'=0$) $\{$  \label{fattwhile}
        \begin{itemize}
            \item If ($a< L_{k^*-1}^{xy}-L_{k}^{xy}$) $k^*\leftarrow \lfloor \frac{(k+1)+k^*}{2} \rfloor $
            \item Else If ($a> L_{k^*}^{xy}-L_{k}^{xy}$) $k^*\leftarrow \lfloor \frac{k^*+K}{2} \rfloor $
            \item Else If ($L_{k^*-1}^{xy}-L_{k}^{xy}\leq a \leq L_{k^*}^{xy}-L_{k}^{xy}$) $a'=L_{k^*-1}^{xy}-L_{k}^{xy}$ $\}$
        \end{itemize}
    \end{enumerate}
    \item $c^{xy}(\tau)=(\tau_{k^*}-\tau)+\frac{a-a'}{v_{k^*}^{xy}}$ $\}$
\end{enumerate}

\subsubsection{Description of $FATT^\tau_{xy}$ Procedure}

\begin{itemize}
    \item Steps 1 and 2: They are identical to the ones of the $ATT$ procedure.
    
    \item Step 3(a): (i) If the procedure is continued to step 3, $a$ is set so as to be equal to the residual distance of arc $<xy>$, if the distance traversed from time instant $\tau$ to $\tau_{k+1}$ is subtracted from $d_{xy}$. The traversal of distance $d_{xy}$ with $\tau$ as the time instant of departure, is equivalent to traversing distance $a$ with $\tau_{k+1}$ as the time instant of departure. 
    
    (ii) For the binary search over $k^*$ (i.e., the index of the time interval $[\tau_{k^*},\tau_{k^*+1})$ where the time instant of arrival lies in), $k^*$ is initialised with the value $\lfloor \frac{(k+1)+K}{2} \rfloor$. 
    
    (iii) Variable $a'$ is set to zero and it is used in steps 3(b) and 4.
    
    \item Step 3(b): Binary search over $k^*$ is performed, to locate the time interval $[\tau_{k^*},\tau_{k^*+1})$ where the time instant of arrival lies in. According to equation \ref{eqnld}, $L_{k^*}^{xy}-L_{k}^{xy}$ is equal to the distance traversed on $<xy>$ from $\tau_{k+1}$ to $\tau_{k^*+1}$. 
    
    \begin{itemize}
        \item If $a< L_{k^*-1}^{xy}-L_{k}^{xy}$, then distance $a$ (with $\tau_{k+1}$ as time instant of departure) will be traversed prior to $\tau_{k^*}$. Index $k^*$ is too large, consequently, $k^*\leftarrow \lfloor \frac{(k+1)+k^*}{2} \rfloor $.  
        \item If $a> L_{k^*}^{xy}-L_{k}^{xy}$, then distance $a$ will be traversed after $\tau_{k^*+1}$. Index $k^*$ is too small, consequently,  $k^*\leftarrow \lfloor \frac{k^*+K}{2} \rfloor $. 
        \item If $L_{k^*-1}^{xy}-L_{k}^{xy}\leq a \leq L_{k^*}^{xy}-L_{k}^{xy}$, index $k^*$ is the ``correct'' one (i.e., for the time instant of arrival $\tau^*$, $\tau_{k^*}\leq \tau^* \leq \tau_{k^*+1}$). Consequently, binary search is terminated and variable $a'$ is set to $L_{k^*-1}^{xy}-L_{k}^{xy}$. The procedure exits the while loop. 
    \end{itemize}
    \item Step 4: Distance $a'$ is the part of $a$ that is traversed from $\tau_{k+1}$ to $\tau_{k^*}$. The residual distance, i.e., $a-a'$, will be traversed in time interval $[\tau_{k^*}, \tau_{k^*+1})$, i.e., with speed $v_{k^*}^{xy}$. Therefore, the traversal time of $<xy>$ with $\tau$ as time instant of departure, is equal to the traversal time up to time instant $\tau_{k^*}$,  plus $\frac{a-a'}{v_{k^*}^{xy}}$, i.e., $c^{xy}(\tau)=(\tau_{k^*}-\tau)+\frac{a-a'}{v_{k^*}^{xy}}$.
\end{itemize}

To further clarify the proposed procedure, note that:

\begin{itemize}
    \item For the quantity $\frac{a-a'}{v_{k^*}^{xy}}$, $0\leq \frac{a-a'}{v_{k^*}^{xy}} \leq l^{xy}_{k^*}$. If $a=L_{k^*-1}^{xy}-L_{k}^{xy}$, then $\frac{a-a'}{v_{k^*}^{xy}}=0$. If $a = L_{k^*}^{xy}-L_{k}^{xy}$, then  $\frac{a-a'}{v_{k^*}^{xy}}=l^{xy}_{k^*}$.  
    \item The way values are assigned to $k^*$, permits the latter to vary from $k+1$ to $K-1$, i.e., being in accordance to the fact that the arrival time (if the procedure has moved to step 3) will lie in a time interval from $[\tau_{k+1}, \tau_{k+1})$ to $[\tau_{K-1}, \tau_{K})$. 
\end{itemize}

The utilisation of the additive effective length in the proposed $FATT$ procedure leads to $O(1)$ time for each iteration of the while loop of step 3(b). Therefore, the complexity of the $FATT$ procedure is of order $O(\log K)$, since $O(\log X)$ is the complexity of binary search over $X$ ordered objects \cite{ahuja}.

If the $FATT$ procedure is used in TD-Dijkstra instead of the $ATT$, the resulting algorithm, called \emph{Fast Time-Dependent-Dijkstra (FTD-Dijkstra)}, has complexity of order $O(n^2+m\log K)$, compared to $O(n^2+mK)$ of TD-Dijkstra.  

The derivation of the additive effective length for arc $<xy>$ can be performed using the following proposed \emph{$AEL^{xy}$ procedure}, applied to arc $<xy>$.

\subsection{$AEL^{xy}$ Procedure}

\begin{enumerate}
    \item $L_0^{xy}=l_0^{xy}$, $i=1$ 
    \item While ($i\leq K-1$)
    \begin{enumerate}
        \item $L_i^{xy}=L_{i-1}^{xy}+l_i^{xy}$
        \item $i\leftarrow i+1$
    \end{enumerate}    
\end{enumerate}

Procedure $AEL$ has $O(K)$ computational complexity; the same stands for space complexity. Therefore, the preprocessing computational complexity of FTD-Dijkstra is $O(mK)$. 

The space complexity is also $O(mK)$, equal to the case where the $AEL$ procedure is not applied, since the network consists of $m$ arcs, each one consisting of $K$ time intervals with their corresponding speeds.

A faster implementation can be performed, if Fibonacci heaps are utilised \cite{fibonacci}, \cite{ahuja}. Under this data structure, the amortised complexity of the selection of the node with the minimum cost (step 2a of TD-Dijkstra) is $O(\log n)$, leading to computational complexity of $O(m\log K + n\log n)$ for FTD-Dijkstra. The latter complexity can be further decreased if specific constraints are valid for the network graph, as detailed below.

\subsection{Special Cases}

Consider the case where the effective lengths are bounded below, i.e., $l_i^{xy}\geq \frac{d^{xy}}{Q}$ $\forall <xy>\in A$ and $\forall i: 0\leq i\leq K-1$. Without loss of generality, consider that $Q$ is integer. Then, in step 3(b) of the $FATT$ procedure, $Q$ consecutive time intervals are enough for the traversal of distance $a$\footnote{with $\tau_{k+1}$ as time instant of departure, as explained previously}, since:

\begin{eqnarray}
     &\sum_{k+1}^{(k+1)+(Q-1)}l_{i}^{xy}\geq\nonumber\\ 
     &\geq ((k+1)+(Q-1)-(k+1)+1)\cdot\frac{d^{xy}}{Q}=&\nonumber\\
     &=Q\cdot\frac{d^{xy}}{Q}=d^{xy}>a& 
\end{eqnarray}

Therefore, the binary search in step 3(b) of the $FATT$ procedure can be constrained in the area from $k+1$ up to $k+1+Q$, instead of up to $K$. This leads to computational complexity of order $O(\log Q)$ for the $FATT$ procedure and, consequently to  $O(m\log Q + n\log n)$ for FTD-Dijkstra. If $Q<K$, then this complexity is lower compared to $O(m\log K + n\log n)$. The resulting algorithm is named \emph{Bounded-FTD-Dijkstra (B-FTD-Dijkstra)}.

If the network has the property that for every arc, every time interval has effective length at least as large as the distance of the arc, i.e., $l^i_{xy}\geq d_{xy}$ $\forall <xy>\in A$ and $\forall i: 0\leq i\leq K-1$, then $Q\leq 1$ and B-FTD-Dijkstra has complexity of $O(m + n\log n)$, i.e., equal to the one of classical Dijkstra's algorithm \cite{dijkstra}, \cite{fibonacci}.

The complexity of the existing TD-Dijkstra algorithm as well as of the proposed FTD-Dijkstra and B-FTD-Dijkstra algorithms, is given in Table \ref{tab1}. It must be stated that all the algorithms of Table \ref{tab1} (existing and proposed) are optimal for the networks under investigation.

\begin{table}
\caption{Complexity of existing and proposed algorithms}
\label{tab1}
\begin{footnotesize}  
\begin{tabular}{llcll}
\hline\noalign{\smallskip}
 & ALGORITHM & PREPROC. & SPACE & QUERY  \\
\noalign{\smallskip}\hline\noalign{\smallskip}
\emph{Exist.} & TD-Dijkstra & - & $O(mK)$ & $O(mK + n^2)$ \\
\emph{Prop.} & FTD-Dijkstra & $O(mK)$ & $O(mK)$ & $O(m\log K + n\log n)$ \\
\emph{Prop.} & B-FTD-Dijkstra ($Q \geq K$) & $O(mK)$ & $O(mK)$ & $O(m\log K + n\log n)$ \\
\emph{Prop.} & B-FTD-Dijkstra ($1< Q < K$) & $O(mK)$ & $O(mK)$ & $O(m\log Q + n\log n)$ \\
\emph{Prop.} & B-FTD-Dijkstra ($Q\leq 1$) & $O(mK)$ & $O(mK)$ & $O(m + n\log n)$ \\
\noalign{\smallskip}\hline
\end{tabular}
\end{footnotesize}  
\end{table}

\section{General Case of Speed as Function of Time Inside the Time Interval} \label{general}

Up to this point, it was assumed (as in \cite{Sung}) that the speed is considered to be \emph{constant inside} every time interval. In this Section, the manipulation of the general case of speed being a \emph{function of time inside} a time interval is proposed. Let this function be $g_k^{xy}(t)$ for time interval $[\tau_k, \tau_{k+1})$ and arc $<xy>$, and $\int g_k^{xy}(t)=G_k^{xy}(t)$. This general case can be handled by generalising the notion of effective length  (equation \ref{effective1}) as follows: 

\begin{equation}
    l_k^{xy}=\int_{\tau_k}^{\tau_{k+1}}g_k^{xy}(t)dt = G_k^{xy}(\tau_{k+1})-G_k^{xy}(\tau_{k})    \label{effective2}
\end{equation}

\subsection{Special Case: Constant Speed Inside the Time Interval} 

The (already investigated) special case of speed being constant in every time interval $[\tau_k, \tau_{k+1})$, equal to the one measured at time instant $\tau_k$ (i.e., $g_k^{xy}(t)=v_k^{xy}$) is derived from equation \ref{effective2} as follows:

\begin{equation}
    l_k^{xy}=\int_{\tau_k}^{\tau_{k+1}}g_k^{xy}(t)dt=v_k^{xy} \int_{\tau_k}^{\tau_{k+1}}dt=v_k^{xy}\cdot(t_{k+1}-t_k)    \label{effective3}
\end{equation}

Let the $FAAT$ procedure for this case be \emph{Constant $FAAT$ ($C-FAAT$)}. Then, $C-FAAT$ is exactly the one presented in Section \ref{fattc}. 

\subsection{Special Case: Linear Speed Inside the Time Interval} 

For the case of speed measurements (as described in Section \ref{def}), it is more natural to assume that the speed in the arbitrary time interval $[\tau_k, \tau_{k+1})$ of arc $<xy>$\footnote{where the speed has been measured at time instants $\tau_k$ and $\tau_{k+1}$ with measured values equal to $v_k^{xy}$ and $v_{k+1}^{xy}$ respectively} is a linear function of time ($g_k^{xy}(t)$), taking the values $g_k^{xy}(\tau_k)=v_k^{xy}$ and $g_k^{xy}(\tau_{k+1})=v_{k+1}^{xy}$, rather than being constant, equal to $v_k^{xy}$ for the entire time interval $[\tau_k, \tau_{k+1})$. To the best of our knowledge, work on this case cannot be found in the literature for the networks under investigation.

Under the aforementioned assumption, $g_k^{xy}(t)$ is derived as follows:

\begin{eqnarray}
    &\frac{g_k^{xy}(t)-v_k^{xy}}{t-\tau_k}=\frac{v_{k+1}^{xy}-v_k^{xy}}{\tau_{k+1}-\tau_k}&  \\ 
    &\Rightarrow  g_k^{xy}(t)-v_k^{xy}=\frac{v_{k+1}^{xy}-v_k^{xy}}{\tau_{k+1}-\tau_k} \cdot(t-\tau_k)& \nonumber \\
&\Rightarrow g_k^{xy}(t)=\frac{v_{k+1}^{xy}-v_k^{xy}}{\tau_{k+1}-\tau_k}t+(v_k^{xy}-\frac{v_{k+1}^{xy}-v_k^{xy}}{\tau_{k+1}-\tau_k}\tau_k)&  \nonumber \\ 
&\Rightarrow g_k^{xy}(t)=\frac{v_{k+1}^{xy}-v_k^{xy}}{\tau_{k+1}-\tau_k}t+
            \frac{v_{k}^{xy}\tau_{k+1}-v_{k+1}^{xy}\tau_{k}}{\tau_{k+1}-\tau_k}&      
\end{eqnarray}

If we set, 

\begin{eqnarray}
&\frac{v_{k+1}^{xy}-v_k^{xy}}{\tau_{k+1}-\tau_k}=R_k^{xy}&\\
&\frac{v_{k}^{xy}\tau_{k+1}-v_{k+1}^{xy}\tau_{k}}{\tau_{k+1}-\tau_k}=S_k^{xy}&
\end{eqnarray}

then, 

\begin{eqnarray}
&g_k^{xy}(t)=R_k^{xy}t+S_k^{xy}&   \label{eqn1}  \\
&\Rightarrow G_k^{xy}(t)=R_k^{xy}\frac{t^2}{2}+S_k^{xy}t&  \label{eqn2}   
\end{eqnarray}

From equations \ref{effective2} and \ref{eqn2}, the effective length is derived as follows: 

\begin{eqnarray}
&l_k^{xy}=R_k^{xy}\frac{\tau_{k+1}^2}{2}+S_k^{xy}\tau_{k+1}-R_k^{xy}\frac{\tau_{k}^2}{2}-S_k^{xy}\tau_{k}& 
\end{eqnarray}

Let the $FAAT$ procedure for the current case be \emph{Linear $FAAT$ ($L-FAAT$)}. Then, $L-FAAT$ is derived from $FAAT$ as follows.

\begin{itemize}
    \item Step 1 remains the same as in $FAAT$.
    \item In step 2 of the $FAAT$, $v_k^{xy}\cdot(\tau_{k+1}-\tau)$ is equal to the distance traversed from $\tau$ to $\tau_{k+1}$. This distance in $L-FAAT$ is given by     
    \begin{eqnarray}
        &\int_{\tau}^{\tau_{k+1}}g_k^{xy}(t)dt=
        R_k^{xy}\frac{\tau_{k+1}^2}{2}+S_k^{xy}\tau_{k+1}-R_k^{xy}\frac{\tau^2}{2}-S_k^{xy}\tau &
    \end{eqnarray}

    \begin{itemize}
        \item If this distance is equal to or larger than $d^{xy}$, then the traversal time $c^{xy}(\tau)$ is given by the solution of equation \ref{a}, as follows (where for simplicity, in the intermediate steps for the derivation of equation \ref{b} from equation \ref{a}, $c^{xy}(\tau)$, $R_k^{xy}$, $S_k^{xy}$ and $d^{xy}$ are written as $c$, $R$, $S$ and $d$ respectively).

\begin{eqnarray}
&\int_{\tau}^{\tau+c^{xy}(\tau)}g_k^{xy}(t)dt=d_k^{xy}& \label{a} \\
&\Rightarrow R\frac{(\tau+c)^2}{2}+S(\tau+c)-R\frac{\tau^2}{2}-S\tau= d& \nonumber\\
&\Rightarrow R(\tau+c)^2+2S(\tau+c)-R\tau^2-2S\tau=2d& \nonumber\\
&\Rightarrow R\tau^2+Rc^2+2R\tau c+2S\tau+2Sc-R\tau^2-2S\tau=2d& \nonumber\\
&\Rightarrow Rc^2+2(R\tau+S)c-2d=0& \nonumber\\
&\Rightarrow c=\frac{-2(R\tau+S)\pm\sqrt{4(R\tau+S)^2+8Rd}}{2R}& \nonumber\\
&\Rightarrow c^{xy}(\tau)=\frac{-(R_k^{xy}\tau+S_k^{xy})\pm\sqrt{(R_k^{xy}\tau+S_k^{xy})^2+2R_k^{xy}d^{xy}}}{R_k^{xy}}& \label{b}
\end{eqnarray}

        \item If this distance is less than $d^{xy}$, the procedure continues to step 3.
    \end{itemize}
    \item In step 3(a)i, $a$ is set to $d^{xy}-R_k^{xy}\frac{\tau_{k+1}^2}{2}+S_k^{xy}\tau_{k+1}-R_k^{xy}\frac{\tau^2}{2}-S_k^{xy}\tau$. Steps 3(a)ii, 3(a)iii and 3(b) remain the same.
    \item In step 4 of the $FAAT$, the quantity  $\frac{a-a'}{v_{k^*}^{xy}}$ is the time needed to traverse distance $a-a'$, where this distance is traversed inside time interval $[\tau_{k^*}, \tau_{k^*+1})$. This time in $L-FAAT$ is given by 

\begin{eqnarray}
 &\int_{\tau_{k^*}}^{\tau_{k^*}+c^{xy}(\tau)}g_{k^*}^{xy}(t)dt=a-a'& \label{c}\\
 &\Rightarrow c^{xy}(\tau_{k^*})=\frac{-(R_{k^*}^{xy}\tau_{k^*}+S_{k^*}^{xy})\pm\sqrt{(R_{k^*}^{xy}\tau_{k^*}+S_{k^*}^{xy})^2+2R_{k^*}^{xy}(a-a')}}{R_{k^*}^{xy}}& \label{d}
\end{eqnarray}

The intermediate steps for derivation of equation \ref{d} from eq. \ref{c} are omitted since they are analogous to derivation of equation \ref{b} from eq. \ref{a}.    
\end{itemize}

Considering the above analysis, the exact steps of the proposed $L-FAAT$ procedure are as follows.

\subsubsection{Steps of the $L-FATT^{xy}(\tau)$ Procedure}

\begin{enumerate}
    \item Locate index $k$ such that $\tau_k \leq \tau < \tau_{k+1}$ \label{fattindex}
    \item If ($R_k^{xy}\frac{\tau_{k+1}^2}{2}+S_k^{xy}\tau_{k+1}-R_k^{xy}\frac{\tau^2}{2}-S_k^{xy}\tau\geq d^{xy}$) 
    
\begin{eqnarray}
 &c^{xy}(\tau)=\frac{-(R_k^{xy}\tau+S_k^{xy})\pm\sqrt{(R_k^{xy}\tau+S_k^{xy})^2+2R_k^{xy}d^{xy}}}{R_k^{xy}}& \nonumber
\end{eqnarray}

    Else $\{$
    \item
    \begin{enumerate}
        \item 
        \begin{enumerate}
            \item $a=d^{xy}-R_k^{xy}\frac{\tau_{k+1}^2}{2}+S_k^{xy}\tau_{k+1}-R_k^{xy}\frac{\tau^2}{2}-S_k^{xy}\tau$
            \item $k^*= \lfloor \frac{(k+1)+K}{2} \rfloor$
            \item $a'=0$
        \end{enumerate} 
        \item While ($a'=0$) $\{$  \label{fattwhile}
        \begin{itemize}
            \item If ($a< L_{k^*-1}^{xy}-L_{k}^{xy}$) $k^*\leftarrow \lfloor \frac{(k+1)+k^*}{2} \rfloor $
            \item Else If ($a> L_{k^*}^{xy}-L_{k}^{xy}$) $k^*\leftarrow \lfloor \frac{k^*+K}{2} \rfloor $
            \item Else If ($L_{k^*-1}^{xy}-L_{k}^{xy}\leq a \leq L_{k^*}^{xy}-L_{k}^{xy}$) $a'=L_{k^*-1}^{xy}-L_{k}^{xy}$ $\}$
        \end{itemize}
    \end{enumerate}
    \item 
    \begin{eqnarray}
    &c^{xy}(\tau_{k^*})=\frac{-(R_{k^*}^{xy}\tau_{k^*}+S_{k^*}^{xy})\pm\sqrt{(R_{k^*}^{xy}\tau_{k^*}+S_{k^*}^{xy})^2+2R_{k^*}^{xy}(a-a')}}{R_{k^*}^{xy}}& \nonumber
    \end{eqnarray}
 $\}$
\end{enumerate}

The computational complexity of the proposed $L-FAAT$ procedure is $O(\log K)$. The routing algorithms for this case are derived from the ones proposed for the case of constant speeds inside the time interval (Section \ref{routing}) just by substituting the $FAAT$ procedure with the $L-FAAT$.

Note that the $FAAT$ procedure can be modified analogously for the case of speed being an alternative function of time (rather than linear) inside the time interval.  

\subsection{Validity of the FIFO Property for the Generalised Case}

In \cite{Sung}, it was proven that the FSM satisfies the FIFO property for the case of constant speeds inside the time interval. In this Section, it is proven that the FIFO property is also valid for the general case of speed being an arbitrary function of time ($g_k^{xy}(t)$) inside the time interval.

Consider that for the arbitrary arc $<xy>$, two vehicles $1$, $2$ depart from node $x$ on time instants $\tau_1$, $\tau_2>\tau_1$ respectively, and arrive at node $y$ on time instants $\tau'_1$, $\tau'_2$ respectively. Obviously,  $\tau'_1 >\tau_1$ and $\tau'_2 >\tau_2$. 

%

To prove that the FIFO property is valid, it must be proven that $\tau'_2 >\tau'_1$. The following cases are possible:
\begin{itemize}
    \item $\tau_2\geq \tau'_1\xRightarrow{\tau'_2 >\tau_2} \tau'_2 >\tau'_1$
    \item $\tau_1< \tau_2< \tau'_1$. The distance $d^{xy}$ is traversed from time instant $\tau_1$ to $\tau'_1$, and it can be split into distances $a$ and $b$ ($a+b=d^{xy}$), where $a$ is the distance traversed from $\tau_1$ to $\tau_2$ and $b$ is the distance traversed from $\tau_2$ to $\tau'_1$. Then,
    
        \begin{eqnarray}
        & \tau_2 >\tau_1 \Rightarrow a>0 \Rightarrow b< d^{xy}& \label{e}
        \end{eqnarray}

From equation \ref{e}, it is concluded that the time from $\tau_2$ to $\tau'_1$ is not enough to traverse the whole arc $<xy>$. Therefore, vehicle $2$ that departs from node $x$ at time instant $\tau_2$ will arrive at node  $y$ at $\tau'_2>\tau'_1$.   
\end{itemize}

The computational complexity of the existing and proposed procedures for the derivation of the road's traversal time according to the time instant of departure, for both the investigated cases of speed having constant value or being a linear function of time inside the time interval, are presented in Table \ref{contrib}.

\begin{table}
\caption{Complexity of existing and proposed procedures}
\label{contrib}
\begin{footnotesize} 
\begin{center} 
\begin{tabular}{ccc}
\hline\noalign{\smallskip}
 Speed inside the time interval & Existing procedure & Proposed procedure   \\
\noalign{\smallskip}\hline\noalign{\smallskip}
\emph{constant value} & $O(K)$ &  $O(\log K)$  \\
\emph{linear function of time} & $-$ & $O(\log K)$ \\
\noalign{\smallskip}\hline
\end{tabular}
\end{center} 
\end{footnotesize}  
\end{table}

\section{Conclusions}    \label{conclusions}

In the  current paper, the problem of shortest path routing in transportation networks (in terms of travelling time) where the speed in several of the network's roads is a function of the time interval, was investigated. More precisely, it was considered that the time horizon was split into time intervals, and the speed depended on the time interval. Fast procedures for the derivation of the road's traversal time according to the time instant of departure have been proposed. For the case of the speed having a constant value inside each time interval (in general, different value for each time interval), the proposed procedure is faster compared to the conventional approaches. Furthermore, for the case of the speed being a linear function of time inside each time interval (in general, different linear function for each time interval), a procedure is also proposed and, to the best of our knowledge, relevant approaches cannot be found in the literature.  
 
The proposed procedures were combined with Dijkstra's algorithm and the resulting algorithms, that are practically applicable and of low complexity, provide optimal shortest path routing in the networks under investigation.

Future research will focus on the combination of the proposed procedures with more advanced shortest path routing algorithms that were designed for static networks, in order to derive their versions for time-dependent networks as modelled in the current paper. Examples of such algorithms are A* \cite{astar}, Contraction Hierarchies \cite{ch}, SHARC routing \cite{sharc}. Although time-dependent versions of these algorithms can be found in the literature (e.g., \cite{Batz} for the time-dependent version of Contraction Hierarchies), these have been designed under the assumption that the traversal time function $f^{xy}(\tau)$ (Section \ref{ttm}) is a  piecewise linear function of $\tau$. However, as shown in Section \ref{nonlinear}, for time-dependent networks as modelled in the current paper, this is not always valid.

\section*{Acknowledgments}

This  work  was  co-funded  by  the  European  Regional  Development  Fund
and  the  Republic  of  Cyprus  through  the  Research  Promotion  Foundation  (Project
New  Infrastructure/Strategic/0308/26).


\begin{thebibliography}{99}
\bibitem{may} A. D. May, {\it Traffic flow fundamentals}, Prentice Hall, 1990.
\bibitem{Sung} K. Sung, M. G. Bell, M. Seong, and S. Park, {\it Shortest paths in a network with time-dependent flow speeds}, European Journal of Operational Research, vol. 121, no. 1, pp. 32–39, 2000.
\bibitem{dijkstra} E. W. Dijkstra, {\it A note on two problems in connexion with graphs}, Numerische Mathematik, vol. 1, pp. 269–271, 1959
\bibitem{Cooke} K. Cooke and E. Halsey, {\it The shortest route through a network with time-dependent internodal transit times}, Journal of Mathematical Analysis and Applications, vol. 14, no. 3, pp. 493–498, 1966.
\bibitem{Delling} D. Delling, {\it Time-dependent SHARC-routing}, Algorithmica, vol. 60, no. 1, pp. 60–94, 2011.
\bibitem{Nannicini} G. Nannicini, D. Delling, D. Schultes, and L. Liberti, {\it Bidirectional A* search on time-dependent road networks}, Networks, vol. 59, no. 2, pp. 240–251, 2012.
\bibitem{Delling2} D. Delling and G. Nannicini, {\it Core routing on dynamic time-dependent road networks}, INFORMS J. on Computing, vol. 24, no. 2, pp. 187–201, Apr. 2012.
\bibitem{Delling3} D. Delling, P. Sanders, D. Schultes, and D. Wagner, {\it Algorithmics of large and complex networks}, J. Lerner, D. Wagner, and K. A. Zweig, Eds. Berlin,
Heidelberg: Springer-Verlag, 2009, ch. Engineering Route Planning Algorithms, pp. 117–139.
\bibitem{Delling4} D. Delling and D. Wagner, {\it Time-dependent route planning}, in Robust and Online Large-Scale Optimization, ser. Lecture Notes in Computer Science, R. Ahuja, R. Mhring, and C. Zaroliagis, Eds. Springer Berlin Heidelberg, 2009, vol. 5868, pp. 207–230.
\bibitem{Ding} B. Ding, J. X. Yu, and L. Qin, {\it Finding time-dependent shortest paths over large graphs}, in Proceedings of the 11th International Conference on Extending Database Technology: Advances in Database Technology, ser. EDBT ’08. New York, NY, USA: ACM, 2008, pp. 205–216.
\bibitem{Batz} G. V. Batz, R. Geisberger, P. Sanders, and C. Vetter, {\it Minimum time-dependent travel times with contraction hierarchies}, J. Exp. Algorithmics, vol. 18, pp. 1.4:1.1–1.4:1.43, Apr. 2013.
\bibitem{Chabini} I. Chabini and S. Lan, {\it Adaptations of the A* algorithm for the computation of fastest paths in deterministic discrete-time dynamic networks}, Trans. Intell. Transport. Sys., vol. 3, no. 1, pp. 60–74, Mar. 2002.
\bibitem{ahuja} R. K. Ahuja, T. L. Magnanti, and J. B. Orlin, {\it Network Flows: Theory, Algorithms, and Applications}, Upper Saddle River, NJ, USA: Prentice-Hall, Inc., 1993.
\bibitem{fibonacci} M. L. Fredman and R. E. Tarjan, {\it Fibonacci heaps and their uses in improved network optimization algorithms}, J. ACM, vol. 34, no. 3, pp. 596–615, Jul. 1987.
\bibitem{astar} P. E. Hart, N. J. Nilsson, and B. Raphael, {\it A formal basis for the heuristic determination of minimum cost paths}, IEEE Transactions on Systems Science and Cybernetics, vol. SSC-4(2), pp. 100–107, 1968.
\bibitem{ch} R. Geisberger, P. Sanders, D. Schultes, and D. Delling, {\it Contraction hierarchies: Faster and simpler hierarchical routing in road networks}, in Proceedings of the 7th International Conference on Experimental Algorithms, ser. WEA’08. Berlin, Heidelberg: Springer-Verlag, 2008, pp. 319–333.
\bibitem{sharc} R. Bauer and D. Delling, {\it SHARC: Fast and robust unidirectional routing}, J. Exp. Algorithmics, vol. 14, pp. 4:2.4–4:2.29, Jan. 2010.
\end{thebibliography}
\end{document}